# On the use of time series to improve signal processing of electrical data

Author : Deparis Jacques[1], Gance Julien[2], Leite Orlando[2]

(1) BRGM, 3 avenue Claude Guillemin, 45100 ORLEANS
(2) IRIS Instrument, 1 Av. Buffon, 45100 Orléans

**Introduction**

Electrical Resistivity Tomography (ERT) has become widely used for engineering and environmental applications in the last couple of decades due to (1) the simplification and automation of resistivity meters and (2) the new generation of inversion software (Loke et al., 2014). Although the initial domain of application remain relevant today, these techniques are increasingly used for deep investigation (for mineral, water and geothermal exploration, Carrier et al., 2019, Lajaunie et al., 2019, Troiano et al., 2019, Yan et al., 2018) and shallow environmental/geotechnical applications (clay detection and identification, Bording et al., 2021). These new applications involve working with complex resistivity, which can be computed in the time domain (TDIP: Time Domain Induce Polarization) or frequency domain (SIP: Spectral Induce Polarization). The applications of induced polarization in electrical resistivity measurements are many and varied. In mineral resource exploration, it can be used to discriminate between different types of conductive minerals (Mao et al, 2016), while in groundwater studies, it provides information on the porosity and permeability of geological formations (Revil et al, 2010), could be discriminate pollution (Colombano 2021, Iravani et al, 2023) or compute pollution water content (Koohbor et al, 2022).

Measuring induced polarization in electrical resistivity studies presents challenges, mainly due to (1) the low signal level measured (some % or ‰ of the primary voltage, Dahlin and Leroux, 2012) and (2) the time of measurement. Concerning first point, the weakness signals leads to a decrease in the signal-to-noise ratio, making accurate measurement of induced polarization more complex. In addition, the presence of high frequency noise and/or non-stationary spontaneous polarization requiring advanced signal-processing methods to improve resolution. For second point, measurement time increases, as chargeability is traditionally measured during an off time (50% duty cycle signal), which has consequence to double the measurement time on the field.

Since the last decade, development of electronic and computer science offer new capability allowing to design new generation of instrument. Indeed, the improvement of the digitization accuracy and speed and the storage memory increase now allow storing the potential and current waveforms at acceptable sampling rate. This makes it possible to develop new algorithms to increase the accuracy of resistivity and chargeability calculations. This new way of doing allows to takes the advantage of the high computing power of the computers that is not present in the resistivity meter itself. In particular, a better estimation Self-Potential (SP) is a crucial point to calculate chargeability (Olsson, P-I, 2018). In addition, the a posteriori processing allows also a double processing both in the time and in the frequency domain. The question therefore arises whether even faster measurements are possible in cases of advance processing.
In this paper, we will present a new processing algorithm using full waveform data in order to better estimate complex resistivity, in time domain or in frequency domain. This processing will make it possible to have a better physical quantification of the subsoil, in terms of geotechnical and hydrogeological parameters for example (Revil et al, 2012 for example). It will also improve measurement performance in order to reduce recording times in the field using commercial Syscal-type devices.

**Method and/or Theory**

A technique often used in addition to the ERT method is the "Induced Polarization" (IP). The type of electrodes array used during IP surveys is the same as for the ERT, given that the IP measurement is taken directly after the corresponding ERT measurement. Induced polarization is a geophysical imaging technique used to identify the electrical chargeability of subsurface materials, such as ore bodies.
Chargeability can be measured using time domain, when the current injection is cut (off time). The potential drops down to a level more or less slowly with the time, and decreases during a relaxation period. The chargeability (in mV/V) value is calculated from the following formula:

$$\text{Eq 1}: M = \frac{1}{V_p(t_a - t_b)} \int_a^b V_t(t) dt$$

Where Vp (in V) is the primary potential measured in DC during the current injection. With the same records, the data can be process in frequency domain in order to estimate complex resistivity (i.e. Amplitude and out-phasing of voltage regarding to current).

IRIS instrument markets the Syscal equipment range since 30 years. They are currently developing a new generation of resistivity meters, called Syscal Terra. Syscal Pro and Syscal Terra are multifunctional multi-electrode systems designed for the investigation, profiling and imaging of resistivity and induced polarization. These two generations are designed to respond to any type of exploration, and can be used to locate faults in fractured aquifers or determine the depth and thickness of aquifers.

Table 1: Parameters of Syscal Pro and Syscal Terra

| Equipment | Syscal Pro | Syscal Terra |
| --- | --- | --- |
| Transmitter | 250 W – 800 V – 2.5 A | 250 W – 800 V – 2.5 A |
| Sampling rate | 100 Hz | 1000 Hz |
| Channel number | 10 | 20 |
| Tx output | Voltage regulation | Voltage and current regulation |
| Resolution | 1 µV | 5 nV |
| IP windows | 20 windows | User define (function of sampling rate) |
| Precision | 0.2% | 0.05% |
| Dynamic range | 21 bits | 31 bits |

Both devices developed by IRIS Instruments support a 250 W – 800 V – 2.5 A transmitter. With Syscal Pro the operator can only regulate the voltage, while with Syscal Terra we can regulate both voltage or current. The dynamic range of the old and new generation is 21 bits and 31 bits, respectively (Table 1). Syscal Pro includes 20 IP windows, while Syscal Terra contains 1 IP window for 10 msec, making it the most powerful of the Syscal line. Moreover, the measurement precision of the older generation is 0.2%, whereas the new equipment is more sensitive and measures the geophysical parameters of the soil with a precision of 0.05%, and the resolution of Syscal Pro is 1 µV, while Syscal Terra has 5 nV.

**Self-potential removal**

Since many years, the self-potential is generally corrected through a simple linear compensation in the resistivity meter. This simple and rapid method has been proven to be reliable for ERT and TDIP survey by Dahlin et al. (2002) and more widely by 20 years of use all over the world for all the possible applications.
Despites its success, Dahlin et al. (2000) pointed out that this compensation was unable to suppress non-linear self-potentials, such as the phenomenon of electrode polarization. This phenomenon occurs when an electrode is used as a receiving electrode after being used as a transmitting electrode. We also developed a new algorithm in order to remove the non-linear and linear self-potential. The new algorithm was validate by making the comparison of apparent resistivity between standard procedure

6[th] Asia Pacific Meeting on Near Surface Geoscience & Engineering
13-15 May 2024 • Tsukuba, Japan

developed by IRIS Instrument (Syscal Pro User Manual), and the new one in time domain and frequency domain. Dataset for this study were acquired in BRGM test site near Marcilly en Vilette in France. For this test, we used 48 electrodes with a spacing of 2 meters. The length of the profile is 94 meters. Acquisition was performed using Wenner configuration. The measured Vp varying from 55 mV to 250 mV. K factor varying from 12m to 126m. The data were acquired using duty cycle configuration in order to compute chargeability during off time. Injection period and off time is equal one seconds (see figure 1).

Figure 1a shows signal with a non-linear self-potential acquire on this field. Figure 1b show the signal with self-potential compensation using traditional algorithm (dashed line) and new algorithm (solid line). In first order (full range voltage measurement, between -200 and 200 mV) both curves are similar. When we decrease the range of voltage measurement (grey curves, right axis), between -2 to 2 mV, we see that the curves computing during the off-time is under 0 for traditional algorithm while the curves computing during off time are centred around 0 for the new algorithm.

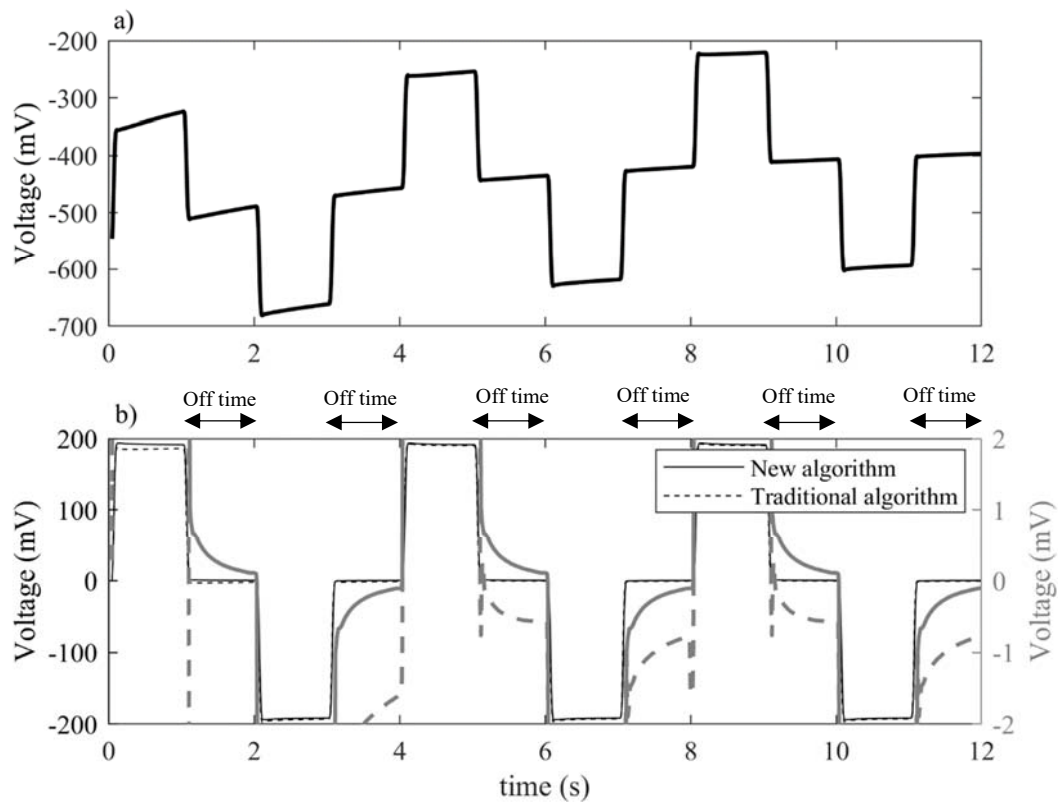

*Figure 1 (a) Example of voltage measurement with non-linear self-potential. (b) Voltage measurement with self-potential correction for traditional algorithm (dashed line) and new algorithm (solid line). The y-axis on the left shows a zoom to see the effect of PS correction on weak signals*

Figure 2 present the voltage measurement obtain during off-time measurement for each stack with traditional (dashed line) and new (solid line) SP compensation. Using traditional algorithm, the decay curves obtained at each stack are "oscillating" around the true one with the pulse polarity. The amplitude of the "oscillations" is decreasing with the measurement time down to zero when the electrode polarization non-linearity becomes null. This phenomenon is sometimes easily detectable when it produces negative chargeability on this level of the pseudo-section but may be sometime also undetectable. Consequently, the accurate TDIP measurement is difficult to perform on the field with multi-electrode systems, and the chargeability section obtained after inversion is always questionable. For the new SP algorithm (solid line), all the line are sur-imposed. The new filter corrects much better the non-linearity of SP. This result is proven by the decay curves obtained at each stacks that are almost all superimposed and that lead to a maximal variation of the total chargeability computed on each.

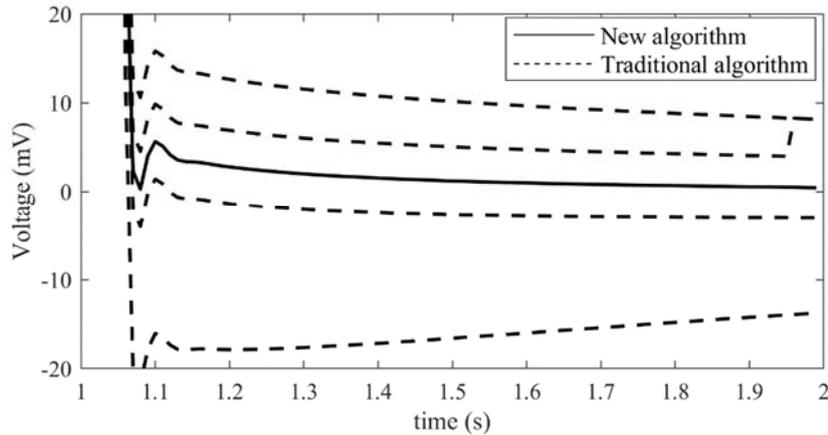

*Figure 2 Stack by stack decay curve with traditional and new SP compensation*

This new SP compensation procedure, which is not based of any assumption, will allow overcoming the electrode polarization effect in TDIP measurements and/or other linear-no linear SP effect. Its use allows obtaining accurate results (Fig. 3), at almost no cost and allows the geophysicist to construct the sequence of acquisition based on signal to noise ratio, vertical and lateral resolution more than on the electrode polarization effect. As a result, acquisition time is reduced.

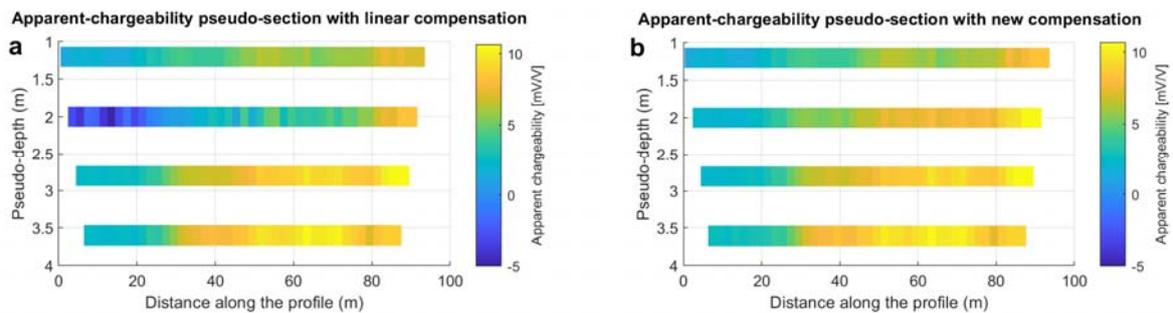

*Figure 3 Comparison of apparent chargeability pseudo-sections measured with a Wenner-Schlumberger configuration (4 levels), PS compensated with a) the linear compensation algorithm and b) the new compensation algorithm. Note negative apparent chargeability on a) have disappeared on b.*

Figure 4 show the comparison between chargeability with the new procedure computed in time domain (a), out-phasing computed in time frequency (b). The results are quite similar for all level of the pseudo section.

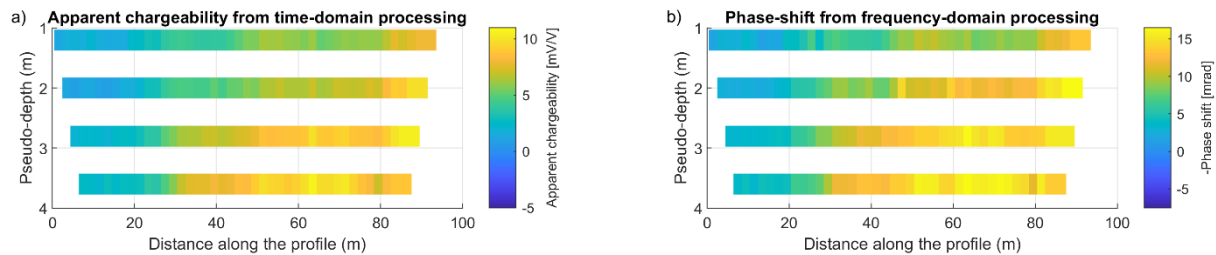

*Figure 4 Comparison of a) apparent chargeability pseudo-sections computed in the time domain and b) the phase shift pseudo-sections computed in the frequency domain. Both processing has carried out using the new PS compensation.*

In order to compare more precisely the induced polarization processing in time and frequency domain, we plotted the correlation between out-phasing and chargeability (Figure 5). A robust fit using matlab© script was computed in order to calculated the regression models. Correlation of both data are good. Indeed, the coefficient of correlation R² is equal to 0.86. This dataset reports that apparent chargeability and apparent phase angle are related by a constant approximately of 1.6, which is in accordance with previous work (Binley, 2015). The obtained equation is the following:

Eq 2: $-\emptyset = 1.6\, M - 0.26$

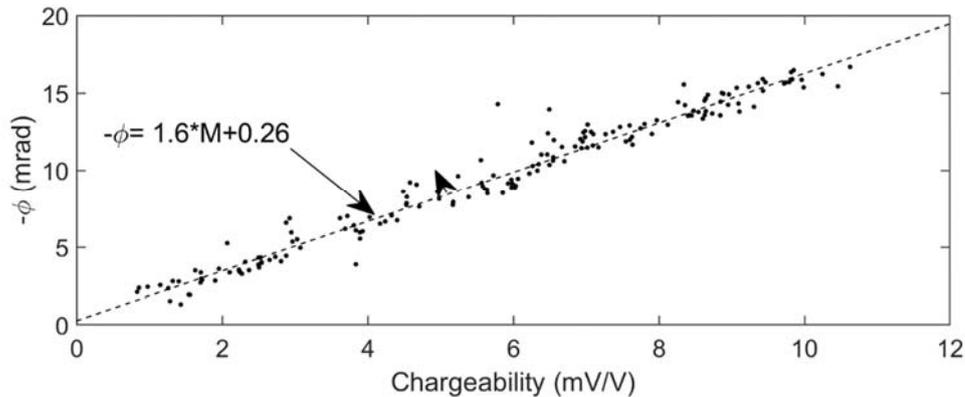

*Figure 5* Correlation between apparent phase shift and apparent chargeability.

**Conclusions**

Since last decade, the resistivity meter offer the capacity to store the full wave data in order to reprocess the signal. In particular, the new generation of Syscal, called Syscal Terra, offer the capacity to store both current and potential signals. We developed a new algorithm for advance processing of complex resistivity in order to increase the signal/noise ratio.

The new algorithm could process data in time and frequency domain. The first step is to better estimate self-potential, especially when the latter is non-linear, in order to remove it. Second step is to compute resistivity and chargeability in time and/or frequency domain. The results show an improvement of the processing, especially for chargeability, when self-potential present non-linear trend. In addition, processing can be realise in time and frequency domain.

This work notice a good linear correlation between calculated out-phasing (in frequency domain) and chargeability (in time domain). It is now possible to estimate chargeability with full cycle signal in frequency domain, which could be have consequence to decrease the time of acquisition (factor of 2 in this case). In addition, future development will used the odd harmonic of square signal signals in order to characterize frequency dependence of complex resistivity. This advanced allow to have a better physical quantification of the subsoil, in terms of geotechnics and hydrogeology for example.

**Acknowledgements**

The authors would like to thank Elodie Courzadet and Laurence Huchet for their help and constructive exchanges, and Iris Instrument for the supply of measurement equipment.